\documentclass[10pt,final,journal]{IEEEtran}
\IEEEoverridecommandlockouts
\usepackage{tikz}
\nonstopmode
\usepackage{cite}
\usepackage{amsmath,amssymb,amsfonts}
\usepackage{graphicx}
\usepackage{textcomp}
\usepackage{xcolor}
\usepackage{hyperref}
\usepackage{diagbox}
\usepackage{cleveref}
\usepackage{algorithm}
\usepackage{algpseudocode}
\usepackage{commath}
\usepackage{optidef}
\usepackage[symbol]{footmisc}
\usepackage{lipsum}
\usepackage{enumitem}
\usepackage{svg}
\usepackage{multirow}

\usepackage[left=1.62cm,right=1.62cm,top=1.7cm,bottom=2.67cm]{geometry}
\IEEEaftertitletext{\vspace{-0.5\baselineskip}}

\makeatletter
\newcommand{\nosemic}{\renewcommand{\@endalgocfline}{\relax}}
\newcommand{\dosemic}{\renewcommand{\@endalgocfline}{\algocf@endline}}

\let\oldnl\nl
\newcommand{\nonl}{\renewcommand{\nl}{\let\nl\oldnl}}
\makeatother

\renewcommand{\arraystretch}{1.25}

\def\BibTeX{{\rm B\kern-.05em{\sc i\kern-.025em b}\kern-.08em
    T\kern-.1667em\lower.7ex\hbox{E}\kern-.125emX}}

\begin{document}

\title{
Machine Learning-Driven Performance Analysis of Compressed Communication in Aerial-RIS Networks for Future 6G Networks
}

\author{ Muhammad~Farhan~Khan, Muhammad~Ahmed~Mohsin, Zeeshan~Alam,
Muhammad~Saad, Muhammad~Waqar \thanks{
(Corresponding Author: $\dag$ Muhammad Farhan Khan )\\
M. Farhan Khan is with the School of Computer Science and Information Technology, University College Cork, Cork, Ireland
(email: Farhan.khan@cs.ucc.ie).

M. Ahmed Mohsin is with the Department of Electrical Engineering, Stanford University, Stanford, CA, USA
(email: muahmed@stanford.edu).

Z. Alam is with Faculty of Computer Science, University of New Brunswick,
NB, Canada
(email: Muhammad.alam@unb.ca).

M.  Saad is with the School of Electrical Engineering and Computer Sciences, SEECS, NUST, Pakistan
(email: msaad.bee20seecs@seecs.edu.pk).

M Waqar is with the Faculty of Arts and Science, Edge Hill University, Lancashire, UK (email:
 m.waqar@uos.acu.uk).

}}





    


\maketitle

\begin{abstract}
In the future 6G and wireless networks, particularly in dense urban environments, bandwidth exhaustion and limited capacity pose significant challenges to enhancing data rates. We introduce a novel system model designed to improve the data rate of users in next-generation multi-cell networks by integrating Unmanned Aerial Vehicle (UAV)-Assisted Reconfigurable Intelligent Surfaces (RIS), Non-Orthogonal Multiple Access (NOMA), and Coordinated Multipoint Transmission (CoMP). Optimally deploying Aerial RIS for higher data rates, employing NOMA to improve spectral efficiency, and utilizing CoMP to mitigate inter-cell interference (ICI), we significantly enhance the overall system capacity and sum rate. Furthermore, we address the challenge of feedback overhead associated with Quantized Phase Shifts (QPS) from the receiver to RIS. The feedback channel is band-limited and cannot support a large overhead of QPS for uplink communication. To ensure seamless transmission, we propose a Machine Learning Autoencoder technique for a compressed communication of QPS from the receiver to RIS, while maintaining high accuracy. Additionally, we investigate the impact of the number of Aerial RIS elements and power allocation ratio for NOMA on the individual data rate of users. Our simulation results demonstrate substantial improvements in spectral efficiency, outage probability, and bandwidth utilization, highlighting the potential of the proposed architecture to enhance network performance.

\end{abstract}


\section{Introduction}
Reconfigurable Intelligent Surfaces (RIS) have garnered significant attention as a solution to enhance the performance of wireless communication networks beyond the fifth generation (B5G) and sixth generation (6G). As urbanization and skyscrapers increase in cities, there is a growing need for an efficient and cost-effective method to enable Non-Line-of-Sight (NLOS) communication, which can be facilitated by RIS. A RIS consists of an array of passive reflective elements on a 2-D meta-surface. Each element can be individually programmed to enhance signal coverage by adjusting both phase and amplitude, enabling efficient and intelligent steering so that the signals are constructively combined at the receiver's end, resulting in higher data rates, improved spectral efficiency, and ubiquitous connectivity while being eco-friendly and cost-efficient. RIS creates a Virtual Line of Sight (VLOS) while passively recycling the signal without any external interference. This makes RIS a crucial enabler for the envisioned smart radio environments of 6G networks. \cite{di_renzo_smart_2020, basar_wireless_2019}.

The advancement of technology has shifted the surge of bandwidth utilization from Orthogonal Multiple Access (OMA) to Non-Orthogonal Multiple Access (NOMA). OMA, which includes TDMA, FDMA, and CDMA, technologies were susceptible to bandwidth exhaustion that could be utilized for other cell users. In contrast, NOMA (Non-Orthogonal Multiple Access) improves bandwidth efficiency and overall network capacity by allowing multiple users to share the same frequency band simultaneously. This requires successive interference cancellation (SIC) to separate user signals based on power levels \cite{jamshed2021unsupervised}. \cite{7842433}, \cite{Dai2018}. By leveraging power domain multiplexing, NOMA can serve multiple users, enhancing throughput and reducing latency \cite{7842433}. The integration of RIS with NOMA systems has been shown to further amplify these advantages by optimizing the signal-to-interference-plus-noise ratio (SINR) and providing more flexible resource allocation \cite{mu_exploiting_2021}.

Integrating CoMP with NOMA techniques is promising for mitigating severe ICI effects in multi-cell NOMA networks and improving spectral efficiency~\cite{Yuan2020, 10464446, Zhang2014, 10462124}. Coordinated multi-point (CoMP) techniques use high-speed front-haul connections and sharing channel state information (CSI) among BSs, that enable coordinated transmissions and enhance overall network performance. However, the challenge of coordination among all BSs still arises due to issues with CSI inaccuracies, synchronization across cells, and increased signal processing requirements.


While the integration of RIS and NOMA has shown promising results in improving spectral efficiency and reducing outage probability \cite{Mu2020, 10847914}, the combination of RIS, NOMA, and CoMP has not been thoroughly investigated, especially with aerial RIS deployments. RIS, mounted on platforms such as unmanned aerial vehicles (UAVs), can provide flexible and dynamic coverage enhancements, adapting to changing network conditions in real-time \cite{Liang2020}. According to the author's best knowledge, no performance analysis exists for Aerial-RIS-aided CoMP-NOMA cellular environments \cite{umer2024deep, 10907822}.

Managing RIS-assisted systems requires phase control for enhanced performance. Many studies have been made for finding optimal phase shifts for different optimization objectives \cite{gao2021outage,
pan2020multicell, umer2024deepreinforcementlearningtrajectory}. Previous studies assumed that these phase shifts are already available at the RIS end in quantized form, but that is not the case. These Quantized Phase Shifts require a feedback channel from the receiver to RIS, which is difficult to obtain due to bandwidth limitations. Bandwidth efficiency or Spectrum efficiency is a crucial requirement for efficient data transmission. A promising approach is to use machine learning-based autoencoders for data compression to achieve this efficiency. Autoencoder is a special type of neural network that can learn to represent the data in a lower dimension and then decode it back to its original form, saving significant bandwidth. The paper investigates using autoencoders to compress the QPS symbols into lower dimensions for a smooth transmission over the feedback channel and decode them at the user end. We can achieve substantial bandwidth reduction while maintaining data integrity by leveraging autoencoders' encoding and decoding capabilities~\cite{mohsin2025vision}. 

Motivated by the literature gap, we present a detailed Machine Learning performance analysis for compressed communication of a system integrating aerial RIS, CoMP, and NOMA. This integrated approach not only enhances performance metrics such as throughput, latency, and coverage but also paves the way for innovative applications in diverse fields ranging from smart cities to autonomous driving \cite{liang_spectrum_2020, liu_intelligent_2021, mohsin2025aienabled6gsemantic}. Our contributions are threefold:
\begin{itemize}
    \item We developed a novel system model for aerial RIS-assisted CoMP-NOMA networks, considering strategic placement of RIS for far-user gain enhancement.
    \item We analyze the impact of various parameters, including RIS elements, and power allocation coefficients, on the system's spectral efficiency and outage probability.
    \item Our system model proposes the far user as a part of two NOMA clusters and our simulation results demonstrate an increase in data rate, network capacity, coverage, and a decrease in outage probabilities.
    \item We compared different autoencoder architectures utilized for compressed communication of encoded QPS bits and observed the reconstruction accuracy of our data from all models.
\end{itemize}

\begin{figure*}[t!]
    \centering
    \includegraphics[width=0.85\textwidth]{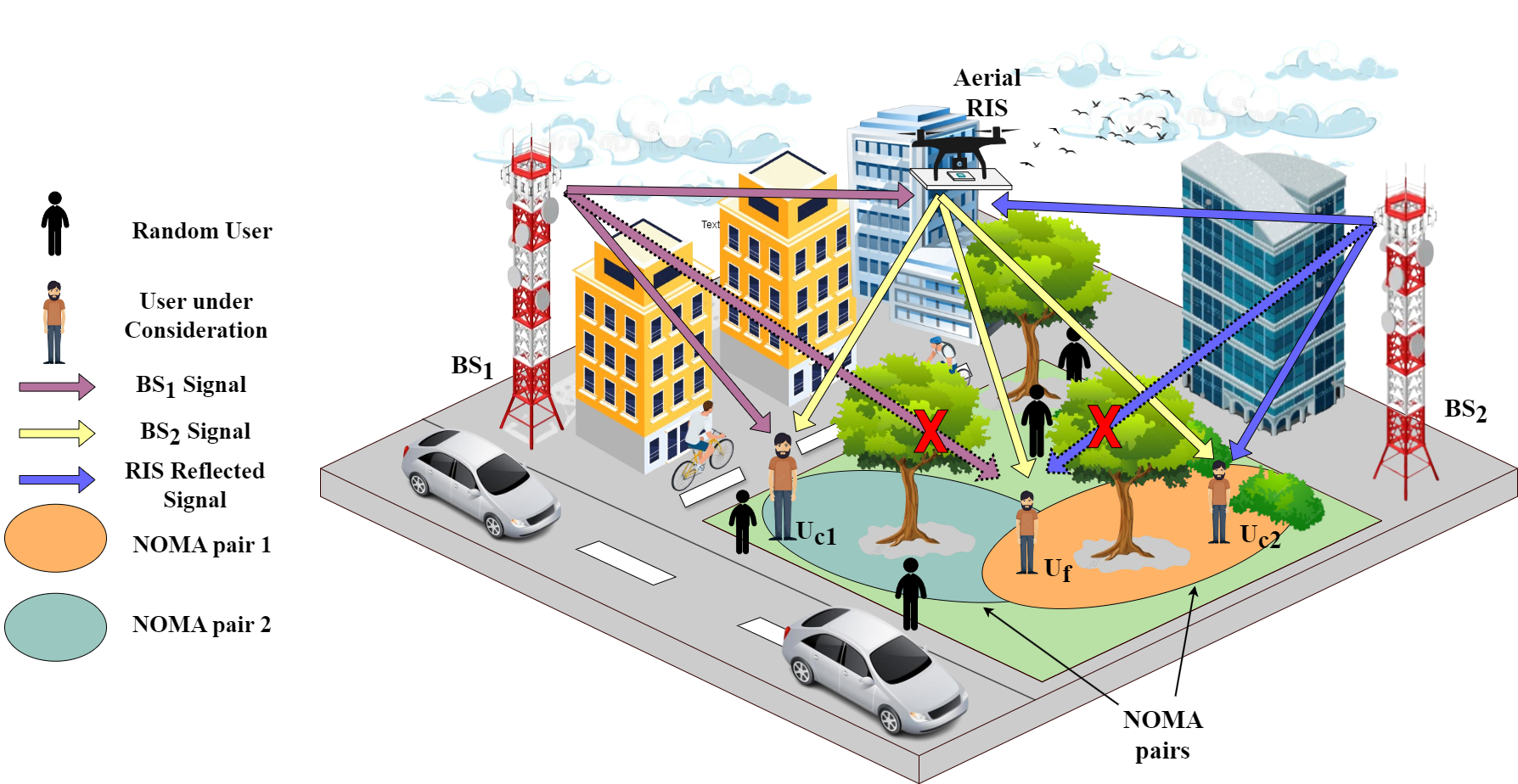}
    \caption{System Model: Power Allocation for UEs (UEn, UEf) with Reconfigurable Intelligent Surface (RIS) Aided Communication}
    \label{fig:sys}
\end{figure*}






\section{System Model and Problem Formulation}
\subsection{System Description}
\begin{figure*}[t]
    \centerline{
        \includegraphics[width=0.9\textwidth]{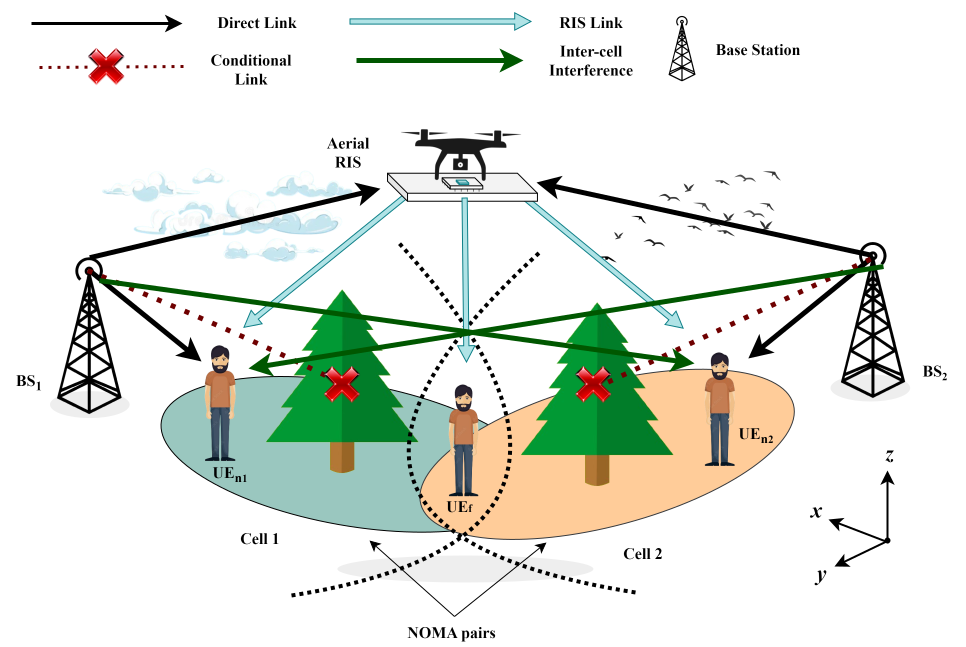}}
    \caption{Simplified System Model}
    \label{fig:simplesys}
\end{figure*}

As described in Fig.~\ref{fig:simplesys}, we consider a two-cell narrow band CoMP-NOMA system with an Aerial-RIS placed strategically at the cell edge of the two cells to facilitate both cell users. The downlink communication system operates with frequency-flat channels. The Aerial RIS consists of $M$ elements. Let the index for the base stations be $\mathcal{C}=\{1,2,\dots,C\}$, $\mathcal{N}=\{1,2,\dots,N\}$ for the two near users and $\mathcal{F}=\{1,2,\dots,F\}$ for the far cell user. Moreover, let $\mathcal{U} = \{\mathcal{N}\,\cup\,\mathcal{F}\}$ be the set of combined users in the system. In Fig.~\ref{fig:sys} a conceptual wireless communication network for a 3D urban environment is visualized, demonstrating the practical use case of our system model. In this urban environment, the direct LOS path is blocked due to obstacles in the environment, and hence aerial RIS provides VLOS. This is a typical example of massive connectivity envisioned for next-gen wireless networks.

To simplify our analysis, we focus on the system model depicted in Fig.~\ref{fig:simplesys}. In this model, each base station is equipped with a single antenna, and the receivers—namely UE$_{c,n}$ and UE$_f$—also have single antennas, establishing a SISO (Single Input Single Output) channel. Furthermore, all $n \in \mathcal{N}$ and $f \in \mathcal{F}$ are located within the reflection region of the aerial RIS, allowing all users in the cellular environment to receive RIS links.

Both base stations use the power domain NOMA technique to transmit signals to their respective NOMA pairs, as shown in Fig.~\ref{fig:simplesys}. Each \( \text{BS}_c \) forms a NOMA pair consisting of UE$_{c,n}$ as the near user and UE$_f$ as the far user. Interestingly, UE$_f$ is part of both NOMA clusters served by both base stations \( \text{BS}_c \). To further enhance the performance of UE$_f$, we adopt Coordinated Multi-point (CoMP), which mitigates inter-cell interference and increases the data rate for this user.

In this research article, we assume perfect channel conditions, implying that perfect Channel State Information (CSI) is available at both base stations \( \text{BS}_c \). Although estimating perfect CSI in real-life scenarios is a highly complex task, recent advancements in research have demonstrated the feasibility of estimating the cascaded wireless channels—from \( \text{BS}_c \) to the Reconfigurable Intelligent Surface (RIS) and from the RIS to the users (UE$_{c,n}$, UE$_f$)—using machine learning and other signal processing techniques ~\cite{he2020channel},~\cite{wei2021channel, 10978440}.

\subsection{Channel Model and RIS Configuration}

In our system model, all communication links account for both large-scale fading and small-scale fading effects. Given our 3D urban environment modeling, the significant distance between base stations \( \text{BS}_c \) and users (UE$_{c,n}$, UE$_f$) results in channels exhibiting Rayleigh fading, characterized by:
\begin{equation}
    \mathbf{h}_{c,u} = \sqrt{\frac{{PL(d}_{0})}{\text{Pd}(d_{c,n})}} \mathbf{w}_{c,n},
\end{equation}
where \( \mathbf{w}_{c,u} \) is a complex Gaussian random variable following a Rayleigh distribution with zero mean and unit variance. \( PL \) denotes the reference path-loss at a reference distance \(d_{0}\) where \(d_{0}\) is taken as 1 meter, and \( \text{Pd}(d_{c,u}) \) represents the large-scale path-loss modeled as \( \text{Pd}(d_{c,u}) = \left(d_{c,u}\right)^{\alpha_{i\rightarrow u}} \), where \( d_{c,u} \) is the distance between BS$_c$ and UE$_u$, and \( \alpha_{i\rightarrow u} \) is the path-loss exponent.

In contrast, the communication link between the Aerial RIS and BS$_1$ manifests a Line of sight (LoS) path~\cite{guo2020intelligent}. As a result, these links are influenced by Rician fading, with their channel coefficient expressed as:
\begin{multline}
    \mathbf{h}_{c,ARIS} = 
    \begin{aligned}
    \sqrt{\frac{{PL(d}_{0})}{\text{Pd}(d_{c,ARIS})}} \left( \sqrt{\frac{\kappa_{c,ARIS}}{\kappa_{c,ARIS} + 1}} \hat{\mathbf{v}}_{c,ARIS} \right. \\
    \left. + \sqrt{\frac{1}{\kappa_{c,ARIS} + 1}} \mathbf{v}_{c,ARIS} \right),
    \end{aligned}
\end{multline}

where $d_{c,ARIS}$ represents the distance between the BS$c$ and the Aerial RIS, $\kappa_{c,ARIS}$ represents the Rician factor, $\hat{\mathbf{v}_{c,ARIS}}$ denotes the deterministic LoS components, and $\mathbf{v}_{c,ARIS}$ denotes the complex Gaussian random variables. Similarly, the communication links between ARIS and receivers, namely UE$_{c,n}$ and UE$_f$— are modeled. The energy splitting (ES) model of the Aerial RIS array is described by its coefficient matrices as follows~\cite{do2021aerial}:
\begin{align}
    \mathbf{\Theta} & = \text{diag}(e^{j \theta_1}, e^{j \theta_2}, \dots, e^{j \theta_M}),
\end{align}
where $\theta_M \in [0, 2 \pi)$, $\forall M \in \mathcal{M} \triangleq \{1, 2,\dots,M\}$.  

\subsection{Problem Formulation}
From the above analysis, the joint optimization problem for maximizing the network sum rate in a single coordinated NOMA cluster aided by the Aerial RIS can be formulated as:
\begin{align}
    \max_{\mathbf{\mathcal{A}}, \mathbf{\Theta}, \textbf{M}_A} \quad & \mathcal{R}_{\text{sum}}=\sum_{c=1}^{2} \mathcal{R}_{c,n} + \mathcal{R}_{f}, \label{eq:opt}             \\
    \textrm{s.t.} \quad                                              & \mathcal{R}_f\geq\text{R}_\text{min}^f, \forall f \in \mathcal{F}, \nonumber                            \\
    \quad                                                            & 
\mathcal{R}_{c,n}\geq\text{R}_\text{min}^{c,n}, \forall c \in \mathcal{C}, n \in \mathcal{N}, \nonumber \\
    \quad                                                            & \gamma_{c,n} + \gamma_f\leq1, \forall c \in \mathcal{C}, n \in \mathcal{N}, f \in \mathcal{F}, \nonumber  \\
    \quad                                                            & \theta_m\in[0, 2\pi), m \in \mathcal{M} \nonumber                                                     \\
    \quad                                         
    &
    \sum_1^c \textbf{M}_A^c \leq M, c \in \mathcal{C}, m \in \mathcal{M}, \nonumber
\end{align}
where $\text{R}_\text{min}^f$ and $\text{R}_\text{min}^{c,n}$ are the minimal achievable rates at UE$_f$ and UE$_{c,n}$, respectively, $\gamma_{c,n}$ and $\gamma_{c,f}$ denote the power allocation (PA) coefficients that BS$_c$ allocates to UE$_{c,n}$ and UE$_f$, respectively, and $\mathbf{\mathcal{A}}$ represents the PA factor of a single NOMA pair within the cluster. $\mathbf{\Theta}$ encompasses all the phase shifts associated with Aerial RIS, while $\textbf{M}_A$ denotes the allocation of Aerial RIS resources among the BSs. Further Rate analysis for far and near user is discussed in the performance analysis section.

The optimal phase shift for each element $m \in \mathcal{M}$, given the CSI being fully estimated at the BSs, can be computed, as referenced ~\cite{wu2019intelligent}, to solve the objective in (\ref{eq:opt}) as follows:
\begin{equation}
    \theta_m = \text{mod}[\arg(\textbf{h}_{i,R}) - \arg(\textbf{h}_{i,R}\cdot \textbf{h}_{R, i}),\,2\pi],
\end{equation}
where the phase of the channels is determined using the argument function $\arg(\cdot)$. To tackle the joint network sum-rate optimization problem (\ref{eq:opt}), we employ fixed empirical optimizations. Additionally, we investigate the effect of allocating Aerial-RIS elements to the BSs by thoroughly examining all possible splitting configurations through an exhaustive iteration process.

Advanced optimization techniques might provide additional improvements. However, this work primarily aims to demonstrate the fundamental benefits achieved by strategically deploying the Aerial RIS and allocating its resources among the BSs in the coordinated cluster.
Due to the limited resolution of the RIS, the phase angle can only assume a finite set of discrete values, typically determined by the quantization process.

\subsection{Machine Learning approach}
The limited resolution of the RIS results in assuming a finite number of discrete values for the phase angle  \( \theta_m \) through quantization. \( \theta_m \) can be represented with quantization levels of \( 2^I \), where \( I \) is the number of quantization bits used. \( \theta_m \) being uniformly quantized, can be expressed in a set of quantized phase shifts (QPS) as \( \{ 0, \frac{2\pi}{2^I}, \dots, \frac{(2^I-1)2\pi}{2^I} \} \). This indicates that Phase Shift Keying (PSK), an appropriate method to map these quantized phases \cite{yu2021convolutional}. In our case, the phase information dataset \(\{x_i\}_{i=1}^m \subset \mathbb{R}^d\), is quantized and encoded to \( I \in \{9\} \) bits, before being transmitted over the communication channel. Quantized bits \( I \) are compressed to \( O \) bits, where \( O \in \{3, 4\} \), resulting in a significant reduction in bandwidth utilization.

Compressed communication was achieved through the implementation of autoencoders. Autoencoders were utilized to encode the QPS bits at the transmitter end and subsequently decode the noisy compressed QPS bits, to retain the original discrete phase shift, aiming to attain efficient communication over a band-limited feedback channel. The models effectively capture the compressed and abstract features of input data, representing them in a latent space and reconstructing the input from the encoded representation. The autoencoder is deployed on the feedback channel, with the encoder located at the user side and the decoder positioned at the RIS. This enables the transmission of compressed QPS information from the user (far or near) to the RIS. The autoencoder can be represented as:

\[
\begin{cases}
\hat{e} = g(X; w_e, b_e) \\
\hat{d} = h(X_e; w_d, b_d)
\end{cases} 
\]
where \(g(\cdot)\) and \(h(\cdot)\) are the encoder and decoder functions, respectively, with each implemented with four different architectures. \(w_e\) and \(b_e\) are the parameters of the encoder, while \(w_d\) and \(b_d\) are parameters of the decoder function and represent the weight matrices and bias vectors of the encoder and decoder networks. We employed four distinct autoencoder architectures for a comparative analysis of our feedback compression mechanism discussed below:

\subsubsection{CNN Autoencoder}

Convolutional Neural Network CNN is a robust architecture for detecting features and patterns from the input data ~\cite{Khan2024a, Khan2023a}. The encoder implementation includes 3 back-to-back Conv + ReLU (a combination of convolution and ReLU activation function), followed by a simple Conv (a simple convolutional layer) layer. Each layer has a typical kernel size of 3 with N neurons, where N is taken as 64. The transmitted QPS are fed to the encoder, represented by I number of bits, with each bit acting as a separate feature, where they undergo a convolution process that is given by:
\begin{equation}
Y_n = (X * w_e)_n= \sum_{k=0}^{K-1} X _k \cdot w_{e,n-k}+ b_e
\end{equation}
where \(X\) is the input vector and \(Y\) is the output feature vector at \(n^{\text{th}}\) position or layer. \(K\) is the total length of the filter.


The ReLU activation function introduces non-linearity which improves the model's ability to learn unique features, particularly the bits associated with QPS in our case. The ReLU function is expressed  as:

\begin{equation}
f(x) = \max(0, X_n)
\end{equation}

The decoder is slightly more complex than the encoder as the reconstruction of the original input requires a comparatively more intricate architecture. Instead of mere Convolutional Layers stacked together, Deep Residual blocks (DRBlocks) are used. In addition to Convolutional layers. The DRBlock is a set of 4 Conv + ReLU layers with an addition of a skip connection that is bridged from the input directly to the output of Convolutional Layers. Overall, the decoder block starts with a Conv layer joined with 2 DRBlocks and a Conv + Sigmoid (a combination of convolutional and sigmoid) as the final layer, with the sigmoid function mapping the output value to [0, 1]. The sigmoid function is given as: 
\begin{equation}
\sigma(x) = \frac{1}{1 + e^{-x}}
\end{equation}

\begin{figure*}[t!]
    \centering
    \includegraphics[width=0.95\textwidth]{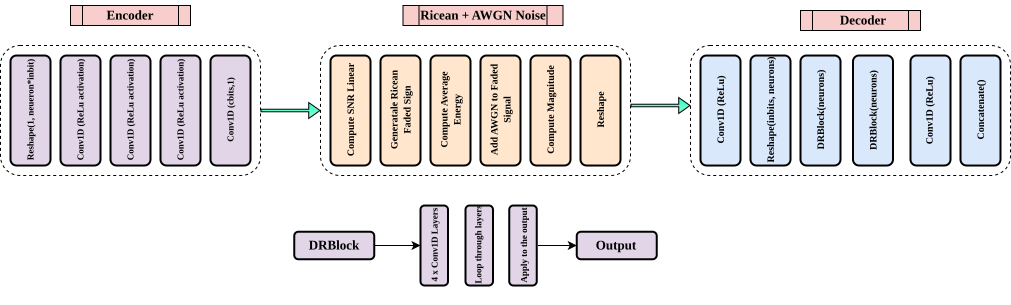}
    \caption{Flowchart illustrating transmission process of bits from encoder to decoder }
    \label{fig:Flowchart}
\end{figure*}

\subsubsection{CNN+Attention Autoencoder}
With the addition of the Attention Mechanism (AM), the performance of the CNN model is further enhanced \cite{niu2021review}. The AM enables the model to autonomously learn to prioritize the most relevant portions of the input. Here, a three-step attention mechanism is utilized. (i) Spatial Attention: This includes a reshaped layer followed by a 1-dimensional convolutional layer with a sigmoid activation function, producing spatial feature maps. An element-wise multiplication is performed with a spatial attention map to highlight significant spatial regions. (ii) Channel Attention: the channel attention permutes and reshapes the spatial output to align the channels, followed by applying a Conv+sigmoid layer. Similarly, an element-wise multiplication is performed with a channel attention map, highlighting the important channels in the data. (iii) Joint-Channel Attention: a 2D Conv + Sigmoid layer is utilized to integrate features across different channels, followed by multiplication with a joint-channel attention map. The final output is the summation of the results of all the mechanisms, producing a more focused and informative representation of the input features.

The encoder is designed with 3 AM layers alternated with three Conv + ReLU layers, each using the same kernel size and number of filters as in CNN architecture. The final output is obtained through a Conv layer with O units for a 1xO representation. The decoder is similar to that of the encoder's design, but instead of Conv+ReLU layers, DRBlocks are utilized, similar to those found in CNN decoder architecture. Similarly, a Conv + Sigmoid layer is employed in the end to regularize the output.

\subsubsection{RNN Autoencoder}
Recurrent Neural Network (RNN) is a deep learning model for sequential data processing. The architecture begins with 3 stacked Simple RNN layers, each with N number of neurons and ReLU activation function. The output is then reshaped to a 1 x NI vector and fed to a final Dense layer for a final compression, mapping the input 1 x I vector to a compressed 1 x O. 
The decoder begins with a dense layer to expand the encoded input into a higher dimension, with a 1 x NI vector representation. The vector is reshaped to form a compatible format with the subsequent RNN layers and then passed through such stack of three Simple Recurrent Neural Network (SimpleRNN) layers with N number of neurons and ReLU activation functions in each layer. In the end, the output of the RNN stack is passed through a time-distributed dense layer, where each time step is processed independently by a dense layer with a single unit and a sigmoid activation function. This final layer produces the reconstructed sequence, with each element in the sequence constrained to the range [0, 1].

\subsubsection{Transformer Autoencoder}

A Transformer Layer is used that comprises of Multi-head Self-attention Mechanism and a Feed-Forward Network (FFN). The multi-head attention mechanism enables the model to capture complex inter-dependencies across different parts of the input sequence, allowing it to focus on various parts of the sequence simultaneously. In our case, 4 multi-heads were used. The Attention Layer is followed by an FFN, comprising a Dense layer with 32 units and a ReLU activation function. FFN applies these non-linear transformations, allowing for more complex representations, followed by dropout and normalization layers. The encoder consists 2 of such Transformer layers, followed by a Reshape and a Dense Layer of O units to map the data to the feature space with the size 1 × O. The decoder has a similar architecture to the encoder, but with an additional dense layer after the multi-head attention layer, to shape the output with 1 x I dimension for the original input reconstruction, so that it can be processed by the final Time-Distributed Dense layer, where each time step is processed independently by a Dense layer with a single unit and a sigmoid activation function.

The encoder's output is augmented with Additive White Gaussian Noise (AWGN) and Rician fading to model the channel interference. $X_e$ is the compressed output signal, scaled with linear signal-to-noise ratio \( \text{SNR}_{\theta_{l}} = 10^{\left(\frac{\text{SNR}_{\text{dB}}}{10}\right)} \) where $\theta_{l}$ shows the linear scale. Accounting for the effects of Rician fading, the normalized Rician fading parameter \( b \) is defined by \( b = \sqrt{\frac{K}{K + 1}} \). K is the Rician factor and represents the degree of dominance of the Line-of-Sight (LoS) component over the non-LoS components in the channel. In a generalized perspective, the faded signal for the feedback channel can be represented as: 
\begin{equation}
Y_{\text{real}} = h_{\text{real}} \cdot X_e
\end{equation}
\begin{equation}
Y_{\text{imag}} = h_{\text{imag}} \cdot X_e
\end{equation}
where \( Y_{\text{real}} \) and \( Y_{\text{imag}} \) are the real and imaginary parts of the received signal, respectively, and \( h_{\text{real}} \) and \( h_{\text{imag}} \) represent the real and imaginary parts of the channel's fading coefficient. The presence of AWGN noise is also considered by calculating its noise power spectral density $\sigma_n^2$, given by:

\begin{equation}
\sigma_n^2 = \frac{E_{\text{avg}}}{\text{SNR}_{\theta_{l}}}
\end{equation}
where $E_{\text{avg}}$ is the average energy and $E_{\text{avg}} = \max\left( \frac{1}{N} \sum_{i=1}^{N} |X_e|^2, 10^{-8} \right)$. Incorporating the AWGN, the signal equation with interference will be expressed as:
\begin{equation}
Z_{\text{real}} = Y_{\text{real}} + N_{o,\text{real}}
\end{equation}
\begin{equation}
Z_{\text{imag}} = Y_{\text{imag}} + N_{o,\text{imag}}
\end{equation}
where \( N_{o,\text{real}} \) is the real part of  noise, and \( N_{o,\text{real}} = \sqrt{\frac{\sigma_n^2}{2}} \cdot \mathcal{N}(0, 1) \), and \( N_{o,\text{imag}} \) is the imaginary part of the noise, and \( N_{o,\text{imag}} = \sqrt{\frac{\sigma_n^2}{2}} \cdot \mathcal{N}(0, 1) \). The noise is modeled as a Gaussian distribution with zero mean and varying variance. The final envelope $|Z|$  of the received signal with the synergy of both AWGN noise and Rician fading will be expressed as:

\begin{equation}
|Z| = \sqrt{Z_{\text{real}}^2 + Z_{\text{imag}}^2}
\end{equation}




\subsection{Training Procedure}
The training procedure was accomplished with end-to-end training for all weights and biases by computing the loss. Each model was trained for 100 epochs, with the rest of the training parameters shown in table \ref{tab:params}. Once the model reaches convergence, additional epochs fail to enhance its accuracy; hence, it was limited to 100. The following equation can characterize the training process:


\[ \Omega = h(q \cdot X_e + N_o), \]
where \( \Omega \) is the output of the autoencoder, \( q \) is the constant channel coefficient, and \( N_o \) is the Additive White Gaussian Noise (AWGN). The loss function utilized during training is the Mean Squared Error (MSE), which is defined as:
\begin{equation}
L = \frac{1}{M} \sum_{m=1}^{M} \|\hat X_m - X_m\|_2^2,
 \end{equation}
where \( \hat X_m \) and \( X_m \) represent the ground truth and predicted output, respectively. M is the total number of training examples. Moreover, the Adam algorithm is selected as the optimizer for updating the model's parameters, and it is formulated as:
\begin{align}
\theta_t &= \theta_{t-1} - \alpha \frac{\hat{m}_t}{\sqrt{\hat{v}_t} + \epsilon} \\
m_t &= \beta_1 m_{t-1} + (1 - \beta_1) g_t \\
v_t &= \beta_2 v_{t-1} + (1 - \beta_2) g_t^2 \\
\hat{m}_t &= \frac{m_t}{1 - \beta_1^t} \\
\hat{v}_t &= \frac{v_t}{1 - \beta_2^t}
\end{align}
where, \( m_t \) is the first-moment estimate (also known as the moving average of the gradients), and \( v_t \) is the second-moment estimate (also known as the moving average of the squared gradients). \( \theta_t \) represents the model parameters at time step \( t \), \( \alpha \) is the learning rate, \( g_t \) is the gradient at time \( t \), and \( \epsilon \) is a small constant added for numerical stability. \( \beta_1 \) and \( \beta_2 \) are exponential decay rates for the moment estimates. Additionally, an exponential decay of 0.99 was used as the learning rate schedule for a gradual decrease in the learning rate during training. The exponential decay is given by:
\begin{equation}
\alpha_t = \alpha_0 \cdot \text{decay\_rate}^{\left\lfloor \frac{t}{\text{decay\_steps}} \right\rfloor}
\end{equation}
where $\alpha_t$ and $\alpha_0$ are the current learning rate and the initial learning rate, respectively.

\begin{table}[t!]
\centering
\caption{Training Parameters for each model }
\label{tab:params}
\begin{tabular}{|c|c|}
\hline
\textbf{Parameter} & \textbf{Value} \\ \hline
Learning Rate & 0.0001 \\ \hline
Decay rate & 0.99 \\ \hline
Batch Size & 500 \\ \hline
Number of Epochs & 100 \\ \hline
Optimizer & Adam \\ \hline
Loss Function & Mean Squared Error \\ \hline
\end{tabular}
\end{table}

\begin{algorithm}
    \caption{Autoencoder for Bit Compression}\label{alg:autoencoder}
    \begin{algorithmic}[1]
        \State \textbf{Input:} \parbox[t]{\dimexpr\linewidth-\algorithmicindent}{Dataset for $UE_{c,n}$ and $UE_f$.}
        
            \State \textbf{Step 1:} \parbox[t]{\dimexpr\linewidth-\algorithmicindent}{Concatenate the dataset for $UE_{c,n}$ and $UE_f$.}
            
            \State \textbf{Step 2:} \parbox[t]{\dimexpr\linewidth-\algorithmicindent}{Bits Compression}
            
            \For{\( \text{epoch} = 1 \) \textbf{to} \( 100 \)}
                \State \parbox[t]{\dimexpr\linewidth-\algorithmicindent}{Pass the input tensor \( X \) to the encoder model with parameters \( \theta_c, \theta_i, \theta_n \).}
                \vspace{0.1em}
                
                \State \parbox[t]{\dimexpr\linewidth-\algorithmicindent}{Pass \( X \) through 4 convolutional layers to get the compressed output \( X_e \) with 4 bits.}
                \vspace{0.1pt}
            \EndFor 

            \State \textbf{Step 3:} Interference Modeling with Fading and Noise.
            \State \parbox[t]{\dimexpr\linewidth-\algorithmicindent}{Pass the output \( X_e \) through Rician noise with \( I=3 \) for dominant LoS to obtain noisy output \( X_r \).}
            \vspace{0.4em}
            \State \textbf{Step 4:} Bits Reconstruction.
            \For{\( \text{epoch} = 1 \) \textbf{to} \( 100\)}
                \State \parbox[t]{\dimexpr\linewidth-\algorithmicindent}{Pass \( X_r \) through the decoder network with parameters \( \theta_c, \theta_i \).}
                \vspace{0.1em}
                \State \parbox[t]{\dimexpr\linewidth-\algorithmicindent}{Pass \( X_r \) through 3 convolutional layers to decode and get the reconstructed output \( X_d \).}
                \vspace{0.1em}
            \EndFor
            
            \State \textbf{Step 5:} Calculate NMSE.
            \For{\( \text{epoch} = 1 \) \textbf{to} \( 100\)}
                \State \parbox[t]{\dimexpr\linewidth-\algorithmicindent}{Calculate Normalized Mean Square Error (NMSE) against Signal to Noise Ratio (SNR).}
                \vspace{0.1em}
                \State \parbox[t]{\dimexpr\linewidth-\algorithmicindent}{Store the calculated NMSE values.}
            \EndFor
        
        \State \textbf{Output:} Reconstructed output \( X_d \) and NMSE values.
            
    \end{algorithmic}
\end{algorithm}

\section{Performance Analysis}
\subsection{Rate Analysis}
We derive the expressions for the achievable rates of the UEs based on the system model. The rate for each user is a function of the power allocation coefficients and the channel conditions, including the effects of the RIS.

For the analysis of the achieved rates for each user in the system model as shown in Fig.~\ref{fig:simplesys}, we first analyze the signal model of the wireless environment. For each $ c  \in \mathcal{C} $ and $ f \in \mathcal{F} $, consider the trio \( \text{UE}_{1,n} \), \( \text{UE}_{2,n} \), \( \text{UE}_f \) as a coordinated NOMA cluster. This setup is essential for assessing the rates achieved and for optimizing system performance within the defined cluster. Each BS$_c$ transmits a superimposed signal containing messages for user equipment in its service area, specifically UE$_{c,n}$ and UE$_f$. This transmission is formulated as ~\cite{saito2013non}:
\begin{equation}
    y_{c}=\sqrt{\gamma_{c,n}P_c}y_{c,n} + \sqrt{\gamma_{c,f}P_c}y_f,
\end{equation}
where $\gamma_{c,n}$ and $\gamma_{c,f}$ denote the power allocation (PA) coefficients and \( P_c \) represents the transmission powers of both \( \text{BS}_1 \) and \( \text{BS}_2 \). Notably, UE$_{c,n}$ enjoys more favorable channel conditions than UE$_f$, establishing it as the primary NOMA user within the pair (UE$_{c,n}, \text{UE}_f$) managed by BS$_c$. By NOMA protocols, UE$_{c,n}$ is expected to successfully detect and decode the message directed to UE$_f$. This operational framework necessitates that $\gamma_{c,n} < 0.5$ and $0.5 < \gamma_{c,f} < 1$ ~\cite{obeed2020user}~\cite{salem2020noma}.
The rate achieved by \( \text{UE}_{1,n} \) is analyzed, which is similar in terms with \( \text{UE}_{2,n} \).

For simplicity and conciseness, the rate achieved by UE$_{1,n}$ is defined only from the set of call center users $\mathcal{N}$, with the understanding that similar steps can be applied to define the rate of UE$_{c,n}$, $\forall c \in \mathcal{C}$ and $n \in \mathcal{N}$. The received signal at UE$_{c,n}$ can be written as:
\begin{equation}
    z_{1,n}=\textbf{h}_{1,n}y_1 + \textbf{h}_{2,n^\prime} y_2 + N_o,
\end{equation}
where $N_o$ represents additive white Gaussian noise (AWGN), specifically $N_o \sim \mathcal{CN}(0, \sigma^2)$. Additionally, $\textbf{h}_{2,n^\prime}$ denotes the channel corresponding to the link between BS$_2$ and UE$_{1,n}$, UE$_{1,n}$ being the cell-center user of BS$_1$, and represents the Inter-Cell Interference (ICI) experienced by UE$_{1,n}$. By employing interference cancellation (SIC) techniques, UE$_{1,n}$ initially decodes the message signal from UE$_f$ (i.e $y_f$), then subtracts it from $z_{1,n}$ to decode its message (i.e $y_{1,n}$). This method allows us to express the signal-to-interference-and-noise ratio (SINR) and the achievable rate at UE$_{1,c}$ for decoding UE$_f$'s message as follows:
\begin{gather}
    \zeta_{1,n\rightarrow f}=\frac{\gamma_{1,f}P_1\abs{\textbf{H}_{1,n}}^2}{\gamma_{1,n}P_1\abs{\textbf{H}_{1,n}}^2 + P_2\abs{\textbf{h}_{2,n^\prime}}^2 +  \sigma^2}, \\
    \mathcal{R}_{1,n\rightarrow f}=\log_2\left(1+\zeta_{1,n\rightarrow f}\right),
\end{gather}
where $\textbf{H}_{1,n}$ represents the combined channel from BS$_1$ to UE$_{1,n}$ and is expressed as $\textbf{H}_{1,n}=\textbf{h}_{1,n}+\textbf{h}_{R, n}^H \mathbf{\Theta_r}\textbf{h}_{1, R}$, represents the combined channel from BS$_1$ to UE$_{1,n}$. Moreover, the Signal-to-Interference-plus-Noise Ratio (SINR) and the achievable data rate of UE$_{1,n}$ for decoding its message can be formulated as follows:
\begin{gather}
    \zeta_{1,n}=\gamma_{1,n}\frac{P_1\abs{\textbf{H}_{1,n}}^2}{P_2\abs{\textbf{h}_{2,n^\prime}}^2 + \sigma^2}, \\
    \mathcal{R}_{1,n} = \log_2\left(1+\zeta_{1,n}\right).
\end{gather}

As a part of two NOMA pairs, UE$_{f}$ receives its signal through the transmission from each BS$_c$, $\forall c \in \mathcal{C}$. Thus, the received signal at UE$_f$ can be expressed as:
\begin{equation}
    z_f = \textbf{H}_{1, f}y_1 + \textbf{H}_{2, f}y_2 + N_0,
\end{equation}
where $\textbf{H}_{1, f}$ and $\textbf{H}_{2, f}$ represent the combined channels from BS$_1$ to UE$_{f}$ and from BS$_2$ to UE$_{f}$, and can be expressed as $\textbf{H}_{1, f}=\textbf{h}_{1, f}+\textbf{h}_{R, f}^H \mathbf{\Theta_t}\textbf{h}_{1, R}$ and $\textbf{H}_{2, f}=\textbf{h}_{2, f}+\textbf{h}_{R, f}^H \mathbf{\Theta_t}\textbf{h}_{2, R}$, respectively. Considering the scenario of non-coherent JT-CoMP, the SINR and achievable rate at UE$_{f}$ can be formulated as~\cite{tanbourgi2014tractable, elhattab2022ris}:
\begin{gather}
    \zeta_{f}=\frac{\gamma_{1,f}P_1\abs{\textbf{H}_{1, f}}^2 + \gamma_{2,f}P_2\abs{\textbf{H}_{2, f}}^2}{\gamma_{1,n}P_1\abs{\textbf{H}_{1, f}}^2 + \gamma_{2,n}P_2\abs{\textbf{H}_{2, f}}^2 + \sigma^2}, \\
    \mathcal{R}_{f}=\log_2\left(1 + \zeta_f\right).
\end{gather}

\begin{figure}[t!]
\centering
\includegraphics[width=\columnwidth]{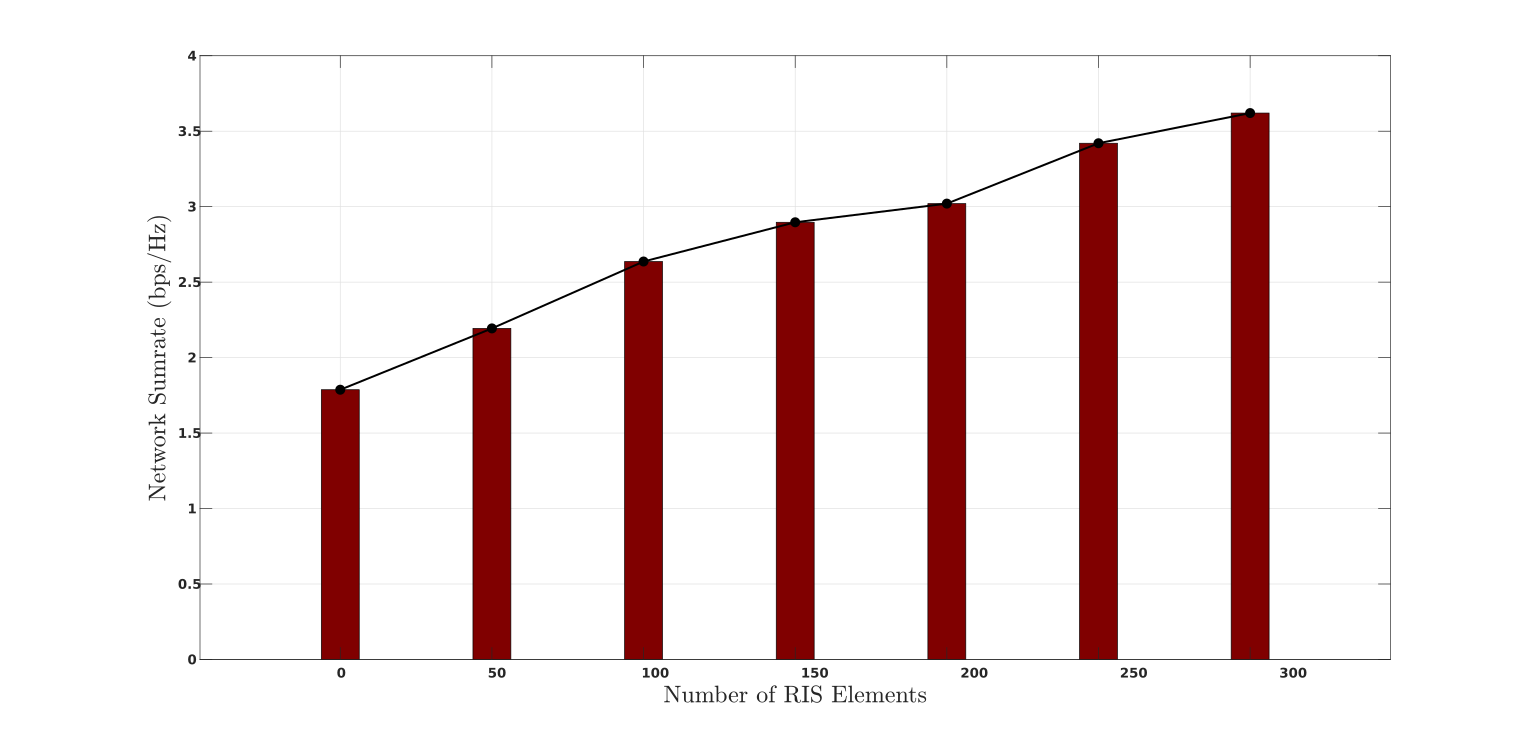}  
\caption{Network sum rate vs. number of RIS elements $M$}
\label{fig:sumrate}  
\end{figure}

\subsection{Outage Probability Analysis}
For further assessing the performance by tactically deploying the Aerial RIS, we analyze the outage probability encountered by cellular users to improve the performance of the system.  According to NOMA principles, for each \( c \in \mathcal{C} \), \( n \in \mathcal{N} \), and \( f \in \mathcal{F} \), if UE$_{c,n}$ fails to decode \( x_f \), or can decode \( x_f \) but not \( x_{i,n} \), an outage event occurs. This outage probability is expressed as
as~\cite{yue2022simultaneously}:
\begin{multline}
    \mathcal{P}_{c,n} = \text{Pr}\,(\zeta_{c,n \rightarrow f} < \zeta_{th_f}) \\ + \text{Pr}\,(\zeta_{c,n \rightarrow f} > \zeta_{th_f}, \zeta_{n} < \zeta_{th_{n}}),
\end{multline}
where $\zeta_{th_f}$ and $\zeta_{th_{n}}$ are the outage thresholds for UE$_{f}$ and UE$_{c,n}$, respectively. Similarly, with regard to UE$_{f}$, an outage occurs when it fails to decode x$_f$, and the corresponding outage probability is expressed as:
\begin{equation}
    \mathcal{P}_{f}=\text{Pr}\,(\gamma_{f}<\gamma_{th_f}).
\end{equation}

\begin{table}[b!]
    \centering
    \setlength{\arrayrulewidth}{0.3pt} 
    \caption{Simulation Parameters}
    \resizebox{0.465\textwidth}{!}{\begin{tabular}{ll}
            \hline
            \textbf{Parameters}                             & \textbf{Values}                      \\ \hline
            Path-loss exponent of BS$_c$-(UE$_n$, RIS) links & $\alpha_{c\rightarrow n}=3.2$          \\\hline
            Path-loss exponent of BS$_c$-UE$_f$ link         & $\alpha_{c\rightarrow f}=4.5$        \\\hline
            Path-loss exponent BS$_c$-RIS links             & $\alpha_{c\rightarrow R}=2.7$          \\\hline
            Path-loss exponent of RIS-UE$_n$ links           & $\alpha_{R\rightarrow c}=3.0$        \\\hline
            Path-loss exponent of RIS-UE$_f$ link            & $\alpha_{R\rightarrow f}=2.7$        \\\hline
            Path-loss exponent of Interfering links         & $\alpha_{c\rightarrow n^\prime} = 4.2$ \\\hline
            Rician factor of RIS-UE$_n$ links                & $\kappa_{R\rightarrow n}=3$ dB       \\\hline
            Rician factor of RIS-UE$_f$ link                 & $\kappa_{R\rightarrow f}=4$ dB       \\ \hline
        \end{tabular}}
    \label{tab:sim}
\end{table}

\section{Results and Discussion}
This section presents our findings on compressed communication in Aerial-RIS networks, enhanced by machine learning. We start by outlining the simulation setup and parameters. We then examine the impact of Aerial RIS elements on urban communication performance and discuss power allocation strategies for improved efficiency. The effectiveness of our methods in maintaining reliable communication is evaluated, followed by a discussion on the trade-offs between spectral and energy efficiency. Finally, we highlight the machine learning results that optimize network performance. These analyses provide insights into the potential and challenges of implementing compressed communication in Aerial RIS networks for 6G and future urban wireless networks.

\subsection{Simulation Setup}
We examine an outdoor scenario where the network operates with a transmission bandwidth \( B = 2.4 \) GHz. The power of the additive white Gaussian noise (AWGN) is calculated as \( \sigma^2 = -174 + 10 \log_{10}(B) \) dBm, considering a noise figure \( NF \) of 12 dB. For simplicity, the transmit powers of both base stations ( BS$_1$ and  BS$_2$) are assumed to be identical, denoted as \( P_t \), ranging from -45 dBm to 0 dBm. 

The power allocation (PA) factors for the near user (UE$_{c,n}$) and the far user (UE$_{f}$) are set to 0.2 and 0.8, respectively. The distances for the far and near users from the base stations are 300 meters and 150 meters, respectively, from their base stations. The key simulation parameters are summarized in Table~\ref{tab:sim}.

\subsection{Impact of Aerial RIS Elements}
The overall network sum rate for users in the assemblage is depicted in Fig.~\ref{fig:sumrate}. With the increase in the percentage of Aerial RIS elements, the network sum rate also increases, as indicated by the upward trend in the graph. The improvement in network performance is due to enhanced data coverage by the RIS. The increase in the RIS elements ensures better signal reflections and transmissions, thereby improving the overall network efficiency and user rates.

\begin{figure}[t!]
\centering
\includegraphics[width=1\columnwidth]{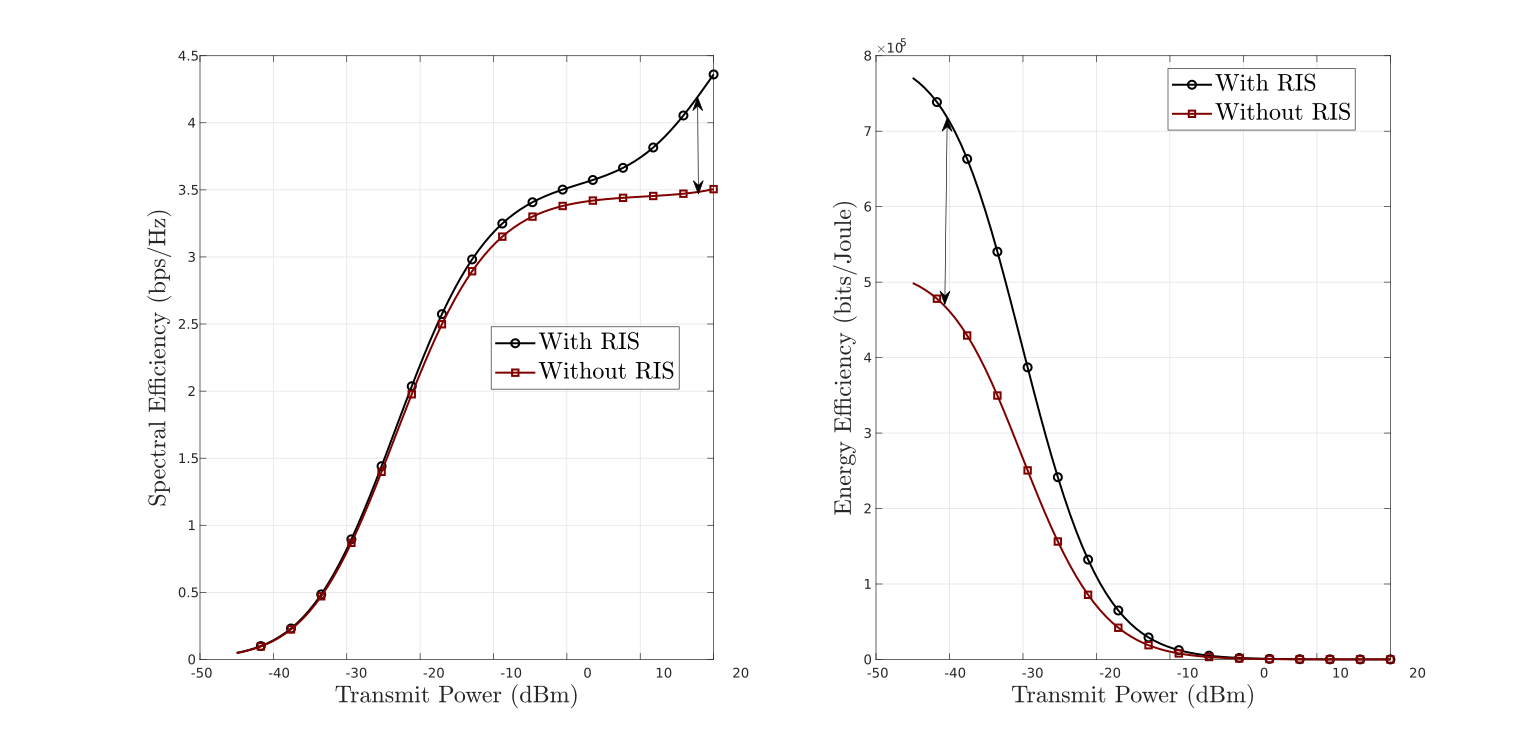}  
\caption{Spectral efficiency and energy efficiency}
\label{fig:rspectral}  
\end{figure}

\begin{figure}[t!]
\centering
\includegraphics[width=1\columnwidth]{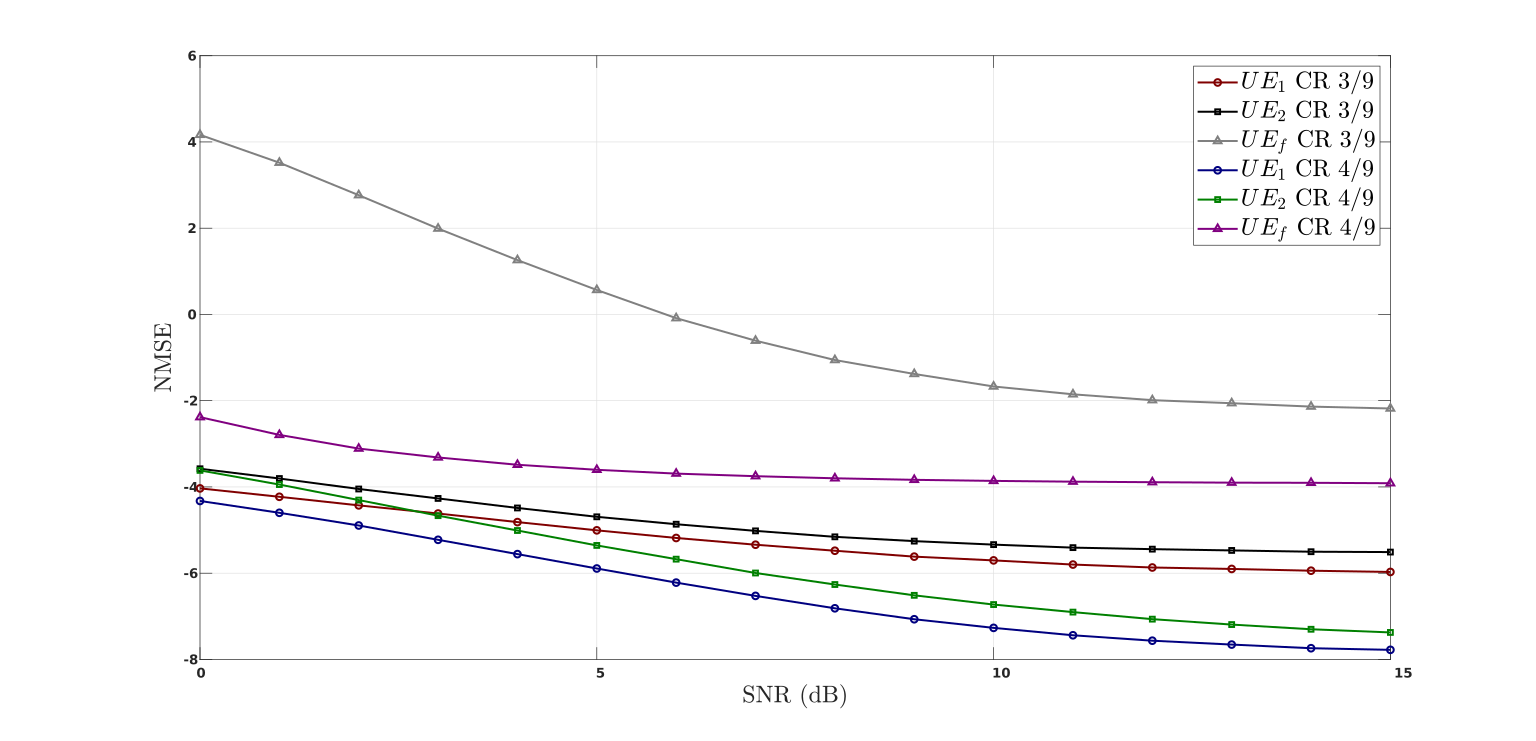}  
\caption{Normalized Mean Square Error (NMSE) of CNN + Attention-based model for $UE_{c,1}$, $UE_{c,2}$, and $UE_{f}$ at CR 3/9 and 4/9.}
\label{fig:nmse49}  
\end{figure}

\subsection{Power Allocation for Coordinated NOMA}
\begin{figure}[t!]
\centering
\includegraphics[width=0.9\columnwidth]{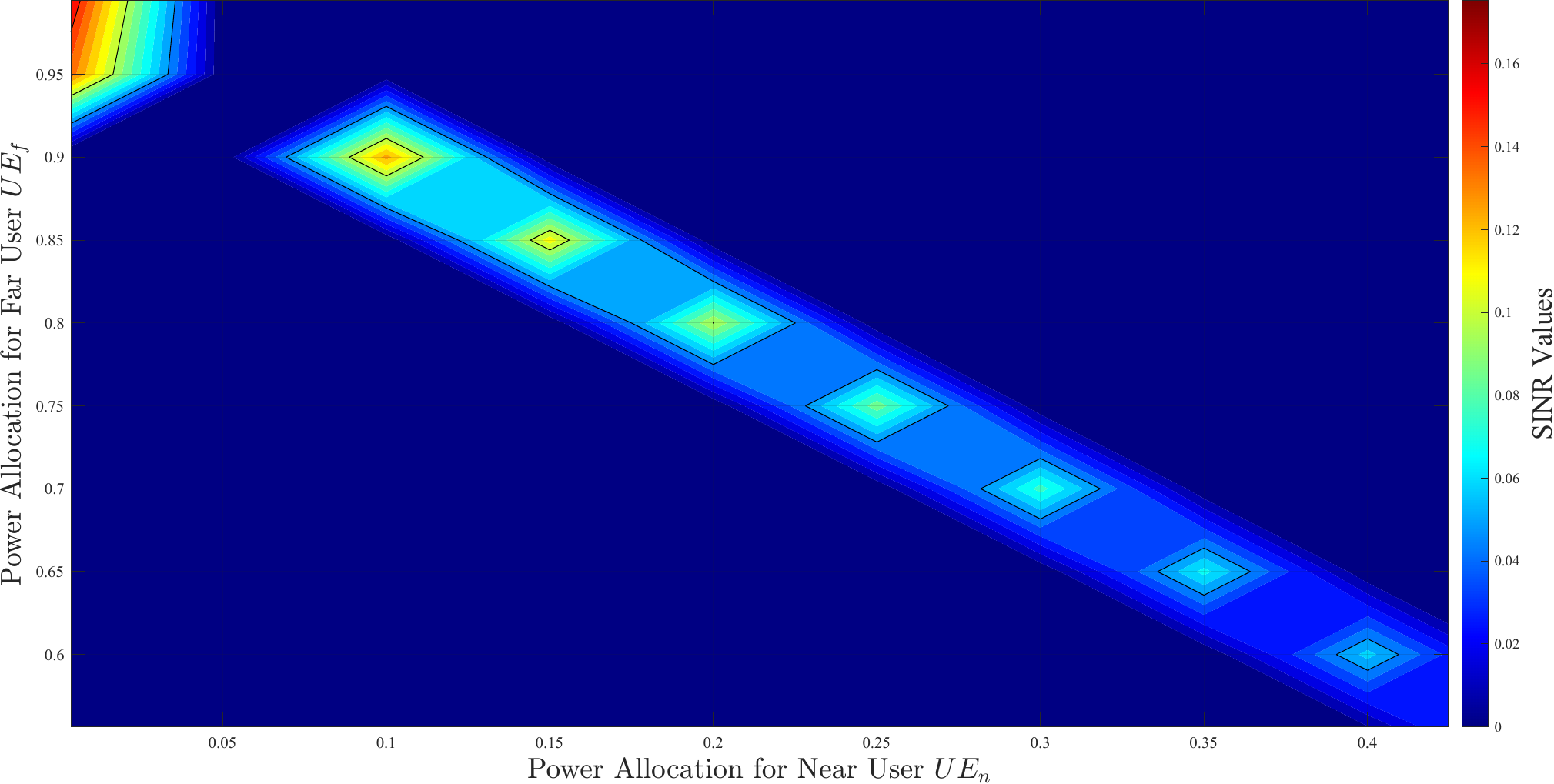}
\caption{Power allocation for $UE_n$ vs. far user $UE_f$}
\label{fig:PowerAllocation}
\end{figure}

Power allocation for users UE$_{c,n}$ and UE$_{f}$ is shown in Fig.~\ref{fig:PowerAllocation}. Within NOMA pairs (i.e., ${\text{UE}_{1,n}, \text{UE}_f}$) and (${\text{UE}_{2,n}, \text{UE}_f}$) as shown in Fig.~\ref{fig:sys}, PA factors are utilized to determine the distribution of transmitted power among users in a paired transmission.

The power allocation factor optimization problem in a NOMA pair can be formulated as follows:

\begin{align}
    \max_{\mathbf{\mathcal{A}}} \quad & \mathcal{R}_{c,n} + \mathcal{R}_{f}, \label{eq:opt2}                                                   \\
    \textrm{s.t.}
    \quad                             & \gamma_{c,n} + \gamma_f\leq1, \forall c \in \mathcal{C}, n \in \mathcal{N}, f \in \mathcal{F}, \nonumber \\
    \quad                             & 0.5 < \gamma_{c,f} < 1, \forall f \in \mathcal{F}. \nonumber
\end{align}

\begin{figure}[t]
\centering
\includegraphics[width=\linewidth]{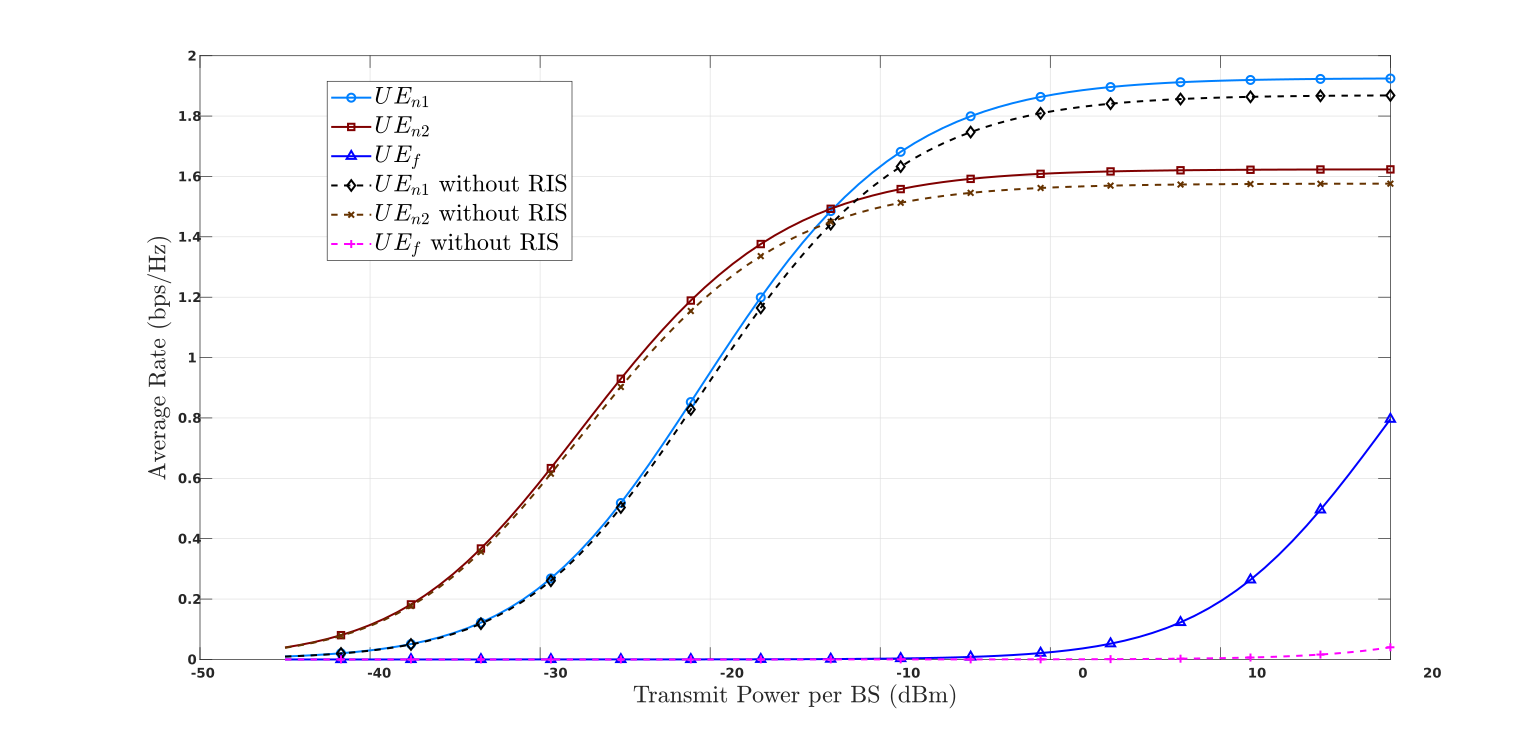}  
\caption{Average rate of $UE_n$,$UE_f$ vs. transmit power per $BS_c$}
\label{fig:rates}  
\end{figure}

 To obtain optimal PA factors for the two NOMA pairs, the methodology proposed in~\cite{fang2016energy} was followed. Additionally, we evaluated the impact of PA factors alongside Aerial RIS enhancements, focused on a configuration with $K=70$ elements. This configuration notably yields the highest network sum rate among all considered cases.

 \begin{figure}[t!]
\centering
\includegraphics[width=\linewidth]{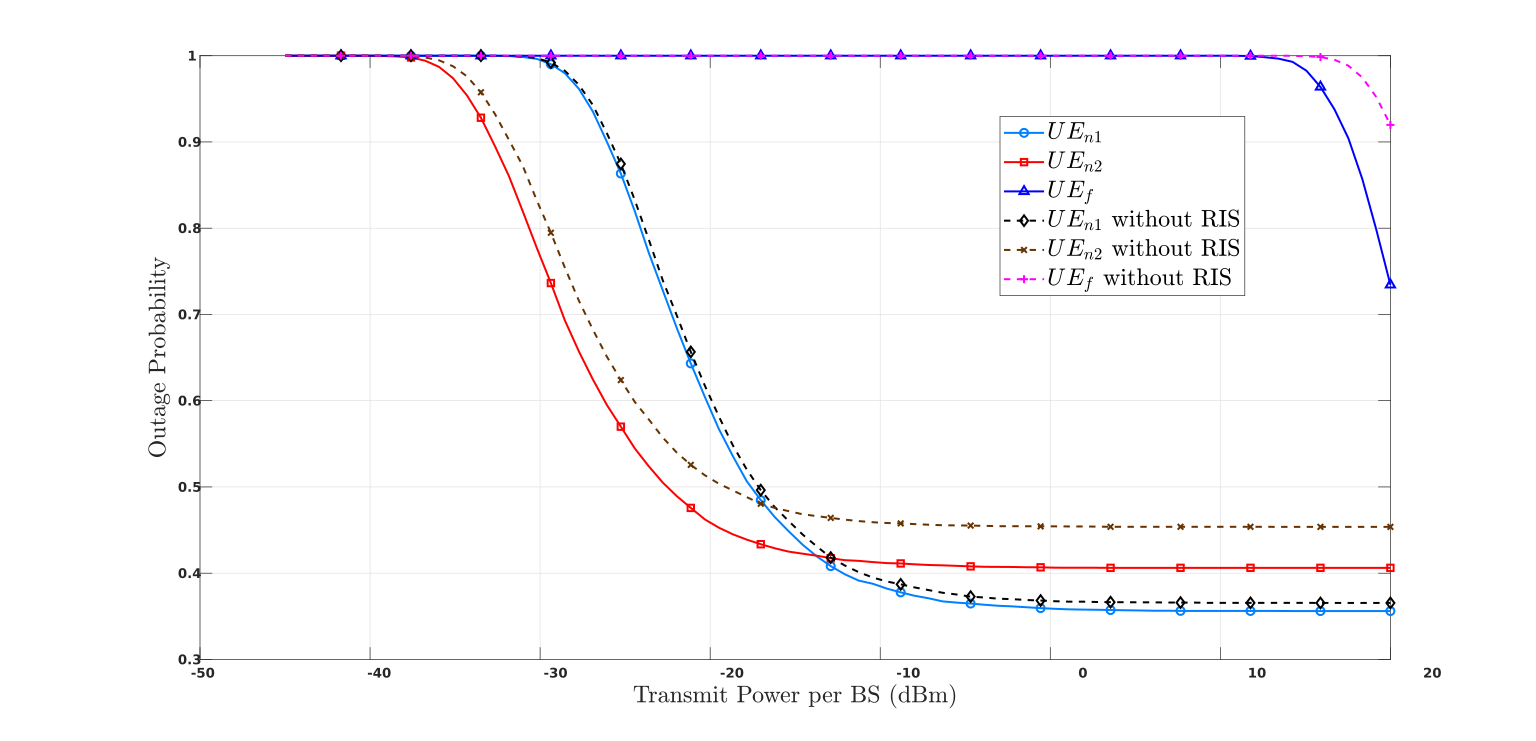}  
\caption{Outage probability of $UE_{c,n}$, $UE_f$, vs. transmit power per $BS_c$}
\label{fig:outage}  
\end{figure}

\begin{table}[b!]
\caption{NMSE values (dB) of Autoencoder architectures for different Compression Ratio (CR)}
\label{tab:comp0}
\centering
\setlength{\tabcolsep}{4pt} 
\renewcommand{\arraystretch}{1.2} 
\begin{tabular}{|l|c|c|c|c|c|c|}
\hline
\multirow{2}{*}{\diagbox[width=2.5cm, height=0.7cm]{\textbf{Models}}{\textbf{CR}}} & \multicolumn{3}{c|}{\textbf{\(\frac{3}{9}\)}} & \multicolumn{3}{c|}{\textbf{\(\frac{4}{9}\)}} \\ \cline{2-7}
& \textbf{$UE_{c,1}$} & \textbf{$UE_{c,2}$} & \textbf{$UE_f$} & \textbf{$UE_{c,1}$} & \textbf{$UE_{c,2}$} & \textbf{$UE_f$} \\ \hline
\textbf{CNN} & -5.393 & -5.076 & -1.868 & -7.755 & -7.164 & -3.551 \\ \hline
\textbf{CNN + Attention} & -5.946 & -5.501 & -2.180 & -7.761 & -7.167 & -3.606 \\ \hline
\textbf{RNN} & -5.596 & -5.042 & -3.059 & -6.880 & -6.313 & -3.110 \\ \hline
\textbf{Transformer} & -2.190 & -2.677 & 4.134 & -2.193 & -2.678 & 4.150 \\ \hline
\end{tabular}
\end{table}

\subsection{Average Rate and Outage Probability}
The average rate and outage probability for each user are depicted in Fig.~\ref{fig:rates} and Fig.~\ref{fig:outage} respectively. The average rate of the communication system, as shown in Fig.~\ref{fig:rates}, increases as the transmit power per BS increases. This is because, increasing the transmit power results in a stronger signal at the receiver, which can help to overcome noise and interference. Fig.~\ref{fig:outage} shows the outage probabilities of $UE_{1,n}$ and $UE_{2,n}$. Both users do not experience any substantial improvements, as their links are already dominated by the near base station $BS_1$ and $BS_2$, respectively. Moreover, due to the strong ICI experienced by $UE_f$ in the non-CoMP configuration, it experiences high outage probabilities, for all transmission power levels. In addition to that, we found that for a given transmit power per BS, the outage probability is lower for UEs with an RIS than for those without one. 

\begin{figure}[t!]
\centering
\includegraphics[width=1\columnwidth]{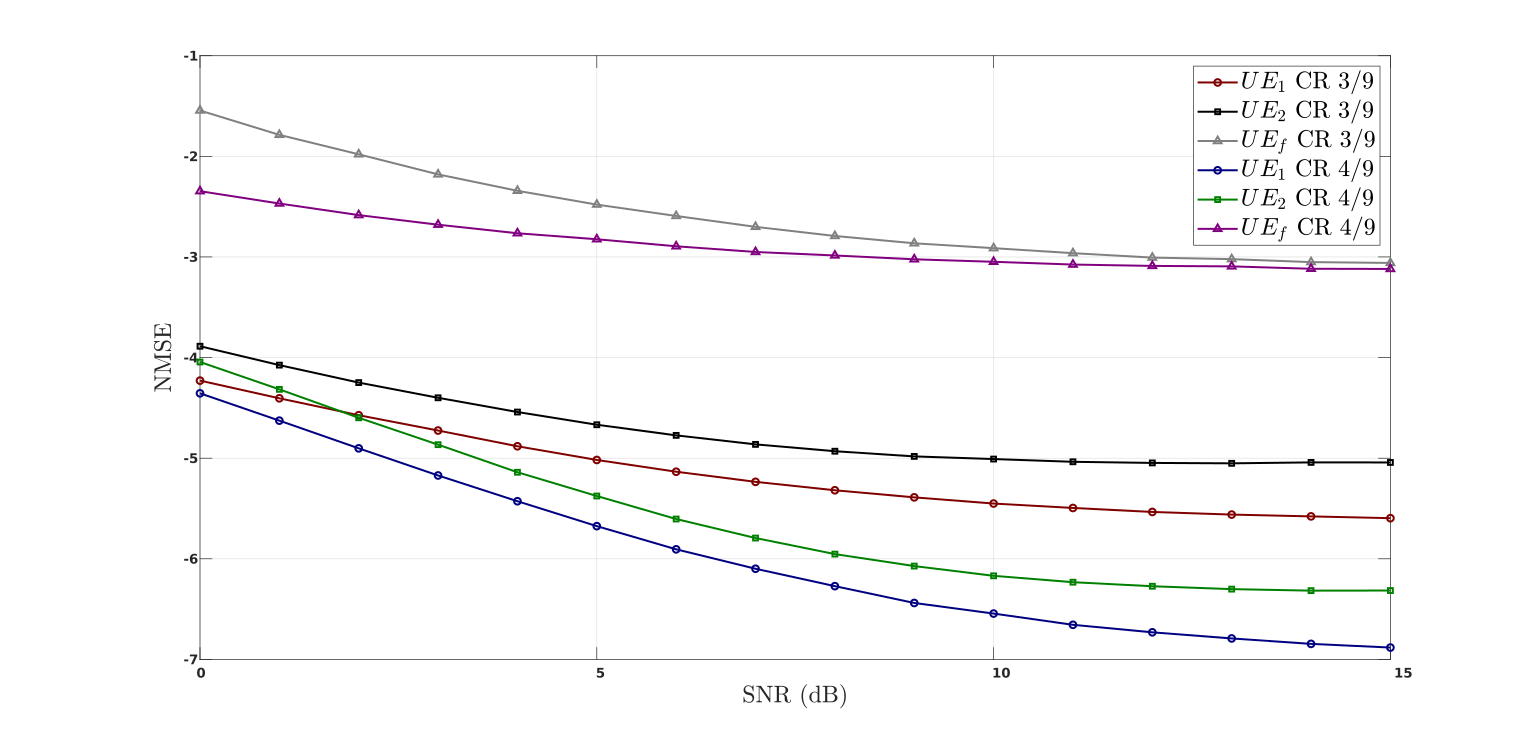}  
\caption{NMSE of RNN-based model for $UE_{c,1}$, $UE_{c,2}$, and $UE_{f}$ at CR 3/9 and 4/9.}
\label{fig:nmse4rnn}  
\end{figure}

\begin{figure}[t!]
\centering
\includegraphics[width=1\columnwidth]{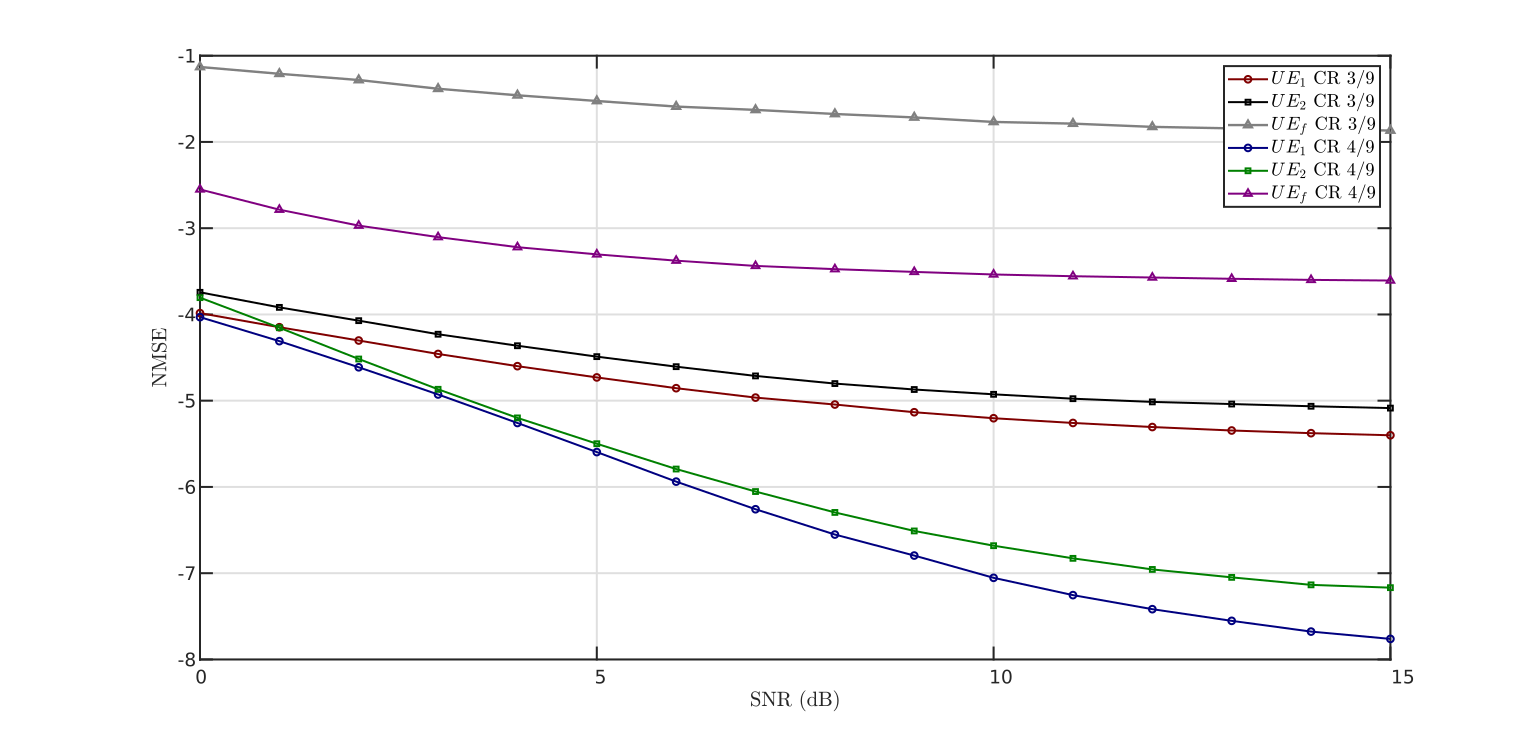}  
\caption{NMSE of CNN-based model for $UE_{c,1}$, $UE_{c,2}$, and $UE_{f}$ at CR 3/9 and 4/9.}
\label{fig:nmse399}  
\end{figure}

\subsection{Balancing Act: Spectral Efficiency vs. Energy Efficiency}
Fig.~\ref{fig:rspectral} portrays the spectral efficiency (SE) and energy efficiency (EE) trade-off as the transmitted power is increased. Spectral efficiency is defined as:    
\begin{equation}
\text{SE} = \log_2(1 + \text{SINR}) \quad \text{(bits/s/Hz)}
\end{equation}
While energy efficiency is given by:
\begin{equation}
\text{EE} = \frac{B \cdot \log_2(1 + \text{SINR})}{P_t} \quad \text{(bits/Joule)}
\end{equation}
where B is the bandwidth and P is the transmitted power of the signal. 
With the inclusion of RIS, the EE significantly improves, particularly at lower transmit power levels and SE at higher transmission power where the RIS gain is more pronounced. This suggests that RIS can effectively mitigate the trade-off between spectral efficiency and energy efficiency, resulting in improved overall performance.

The results indicate that the integration of aerial RIS with CoMP-NOMA significantly enhances network performance. The RIS units effectively mitigate interference and improve signal quality, leading to higher spectral efficiency and lower outage probability.

\begin{figure}[t!]
\centering
\includegraphics[width=1\columnwidth]{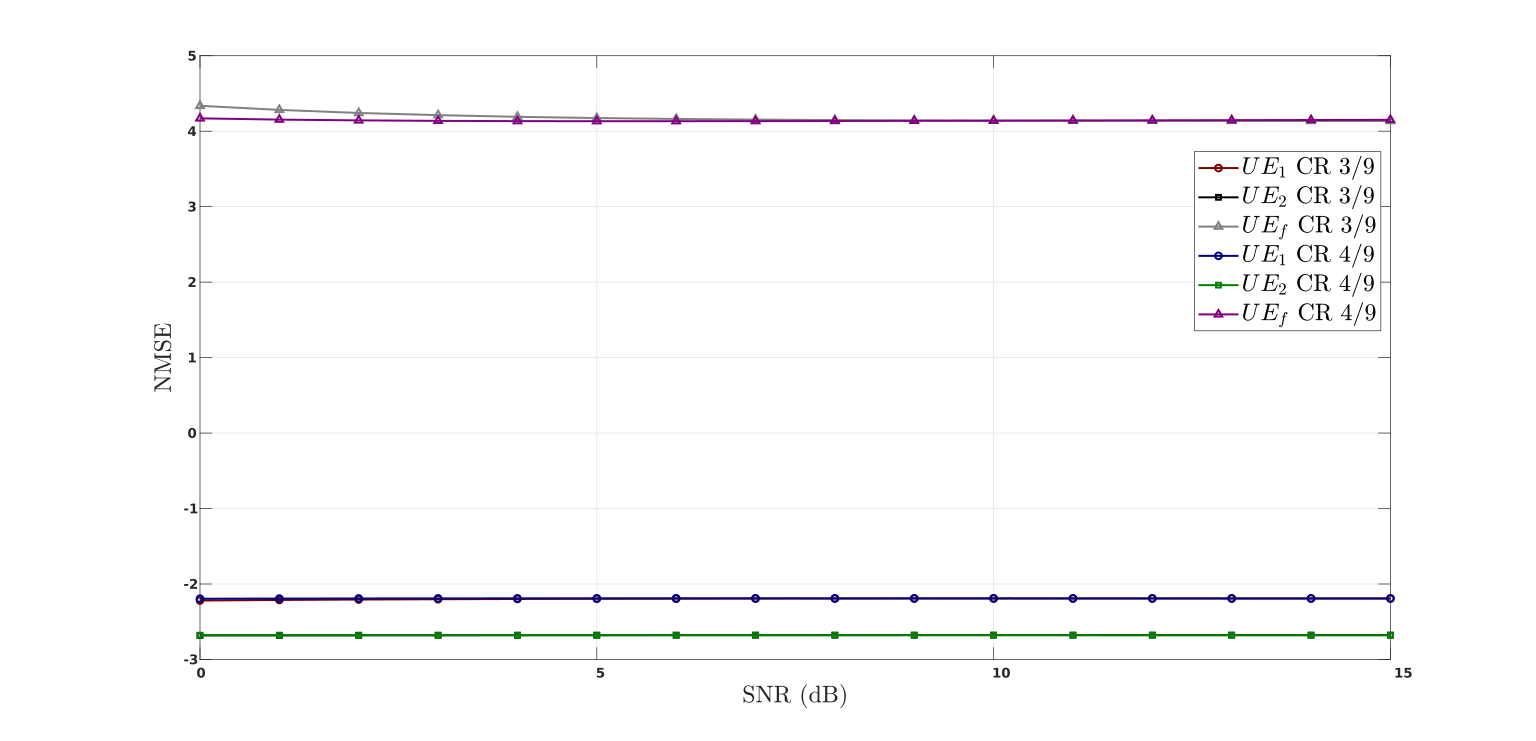}
\caption{NMSE of Transformer-based model for $UE_{c,1}$, $UE_{c,2}$, and $UE_{f}$ at CR 3/9 and 4/9.}
\label{fig:nmse4tran_trans}  
\end{figure}

\begin{figure}[t!]
\centering
\includegraphics[width=1\columnwidth]{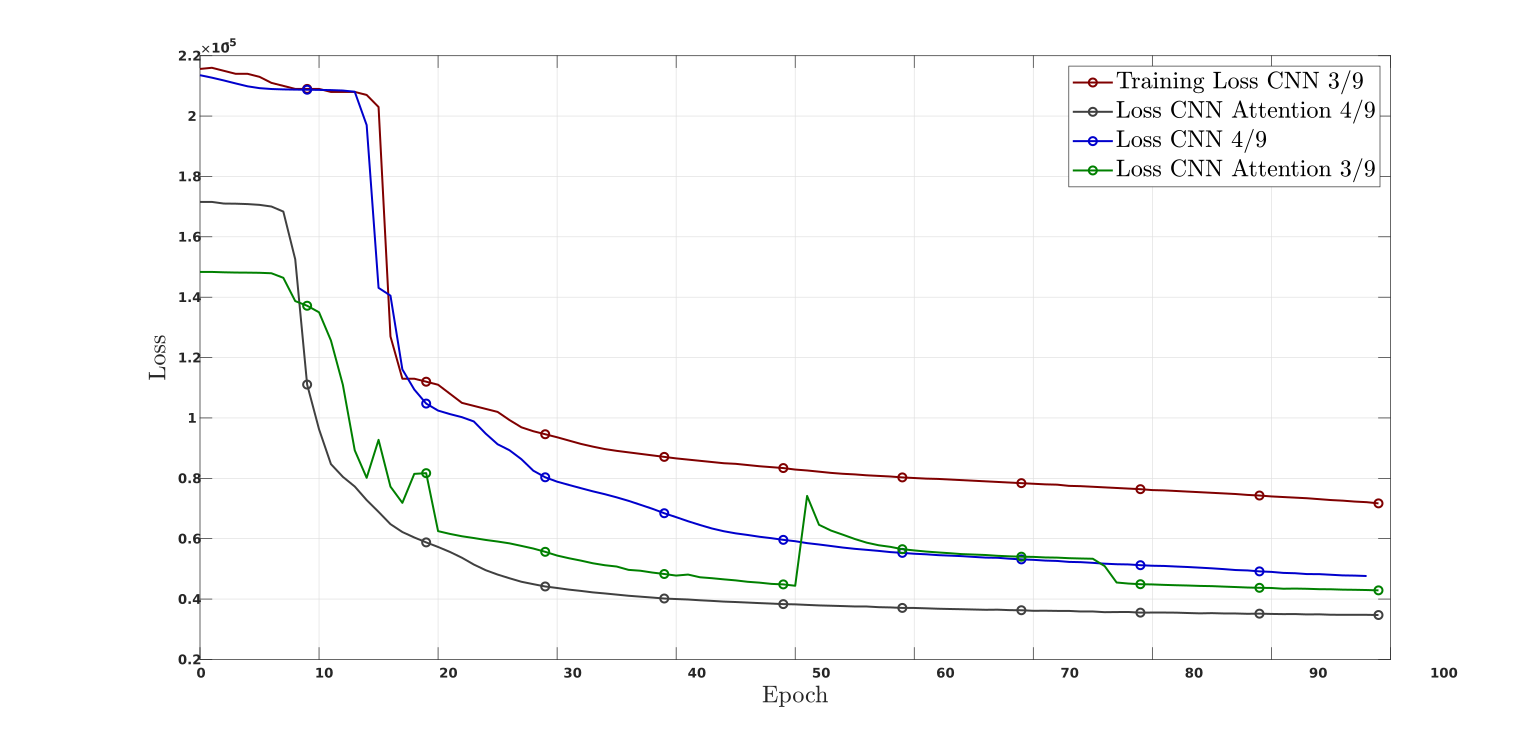}  
\caption{Training loss of CNN and CNN + attention models at CR 3/9 and 4/9}
\label{fig:loss499}  
\end{figure}

\subsection{Machine Learning Results}
The results for the autoencoder were evaluated based on Normalized Mean Square Error (NMSE) as shown in \ref{tab:comp0}, given as:

\begin{equation}
\text{NMSE} = \frac{\| \hat{\Theta}_m - \Theta_m\|_2^2}{\| \hat {\Theta}_m \|_2^2}
\end{equation}
where \(\Theta_m\) is the true value, \(\hat{\Theta}_m\) is the estimated value, \(\| \cdot \|_2\) denotes the Euclidean norm, and the Training Loss of the models employed for our feedback channel. The NMSE showcases the decrease in the model's loss when the SNR associated with each bit increases. For each user, $UE_f$ and $UE_{c,n}$, Compression Ratio (CR) i.e. the ratio of compressed to the original bits \(\frac{O}{I}\), of \(\frac{3}{9}\) and \(\frac{4}{9}\) were used to compute the NMSE with a scale of Signal to Noise Ratio ranging from 0dB to 15dB, whereas the training loss was computed for 100 epochs as mentioned above.

Fig.~\ref{fig:nmse399} shows the performance of CNNs in terms of NMSE for all users accounted in the system model, with both CR. As for the trends of the graph, the NMSE gradually decreases with the increase in SNR. The NMSE for $UE_f$ is significantly greater than that of $UE_{c,n}$ in both scenarios, which is expected due to the larger signal coverage distance for far users compared to near users. The \(\frac{4}{9}\) CR has a better NMSE since the compressed output was associated with more original bits than that of CR \(\frac{3}{9}\). Overall, the CNNs showcase fine results owing to their high capability of learning important features from data.

The results of NMSE of CNN+Attention architecture are displayed in Fig.~\ref{fig:nmse49}. With the addition of AM, CNN's performance is enhanced adequately. The final values of NMSE decayed to a significant value relative to the CNNs, making them more reliable to deploy. The trends for curves are similar to that of CNNs, as NMSE decreases with the increase in SNR until it stabilizes at the end.

\begin{figure}[t!]
\centering
\includegraphics[width=1\columnwidth]{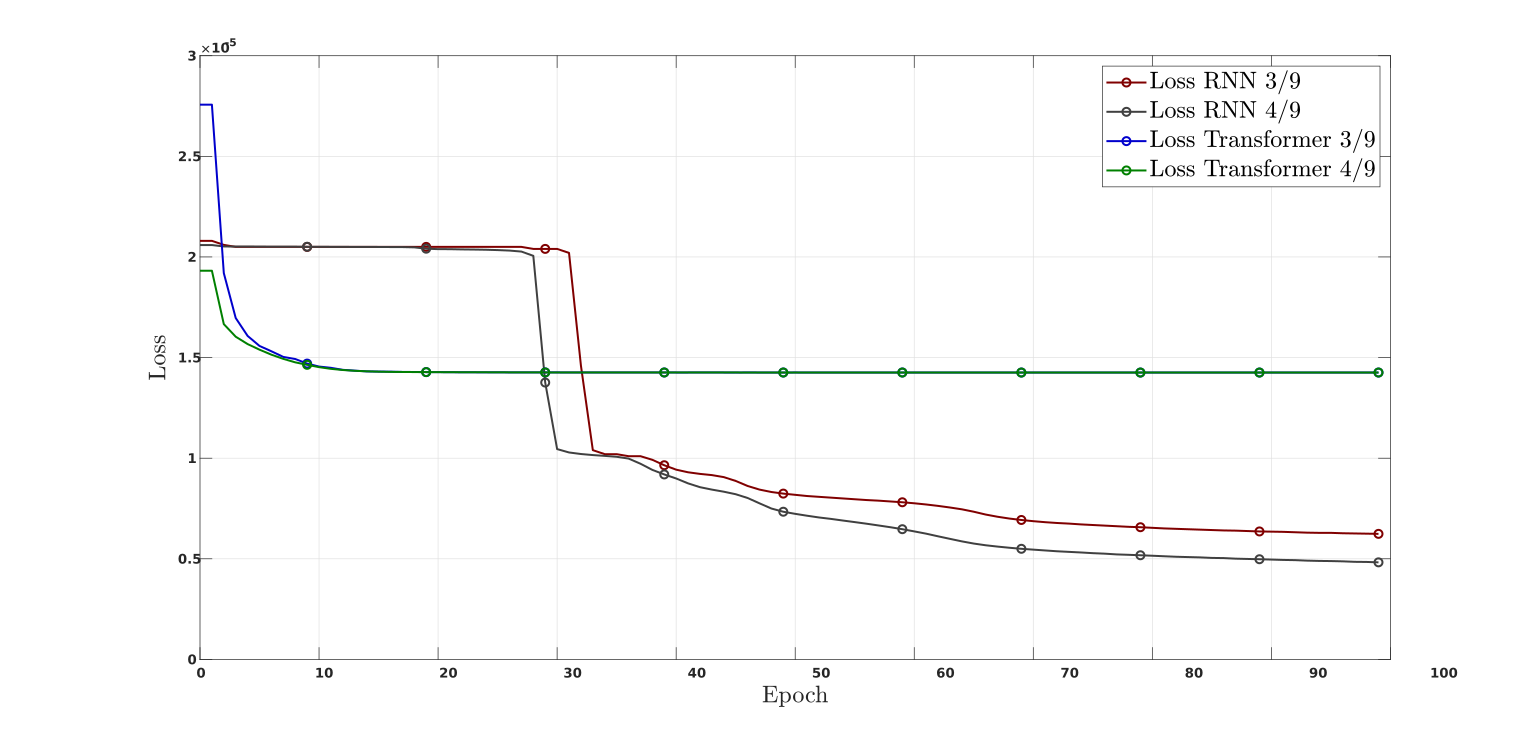}  
\caption{Training loss of RNN and Transformer models at CR 3/9 and 4/9}
\label{fig:nmse4tran_loss}  
\end{figure}

\begin{table}[t!]
\caption{Loss values of Autoencoder architectures for different CR}
\label{tab:comp}
    \centering
    \setlength{\tabcolsep}{3pt} 
    \begin{tabular}{|c|c|c|c|c|}
    \hline
    \diagbox[width=1.8 cm, height=0.7cm]{\textbf{CR}}{\textbf{Models}} & \textbf{CNN} & 
    \textbf{CNN + Attention} & \textbf{RNN} & \textbf{Transformer} \\ \hline
    \textbf{\(\frac{3}{9}\)} & 0.00075 \% & 0.00040 \% & 0.00070 \% & 0.00143 \% \\ \hline
    \textbf{\(\frac{4}{9}\)} & 0.00046 \% & 0.00037 \% & 0.00050 \% & 0.00141 \% \\ \hline
\end{tabular}
\end{table}

Fig.~\ref{fig:nmse4rnn} and Fig.~\ref{fig:nmse4tran_trans} show the performance of RNNs and Transformers in terms of NMSE respectively. RNNs somewhat showcase decent results, having similar trends as that of CNN and CNN+Attention. On the other hand, transformers underperform in this scenario, displaying negligible change in NMSE values over the SNR range. While Transformers can understand complex connections, they aren't as naturally suited to recognize spatial patterns as efficiently as CNNs.

The training loss for CNN and CNN+Attention, for both 3/9 and 4/9 CR is shown in Fig.~\ref{fig:loss499}. The training loss decreases and eventually stabilizes at the end. The loss for CNN+Attention is less than CNNs for both CR as shown in table \ref{tab:comp}, indicating better reconstruction of QPS bits at the receiver end. Similarly, Fig.~\ref{fig:nmse4tran_loss} displays the loss for RNNs and Transformers, respectively.

\begin{figure}[t!]
\centering
\includegraphics[width=1\columnwidth]{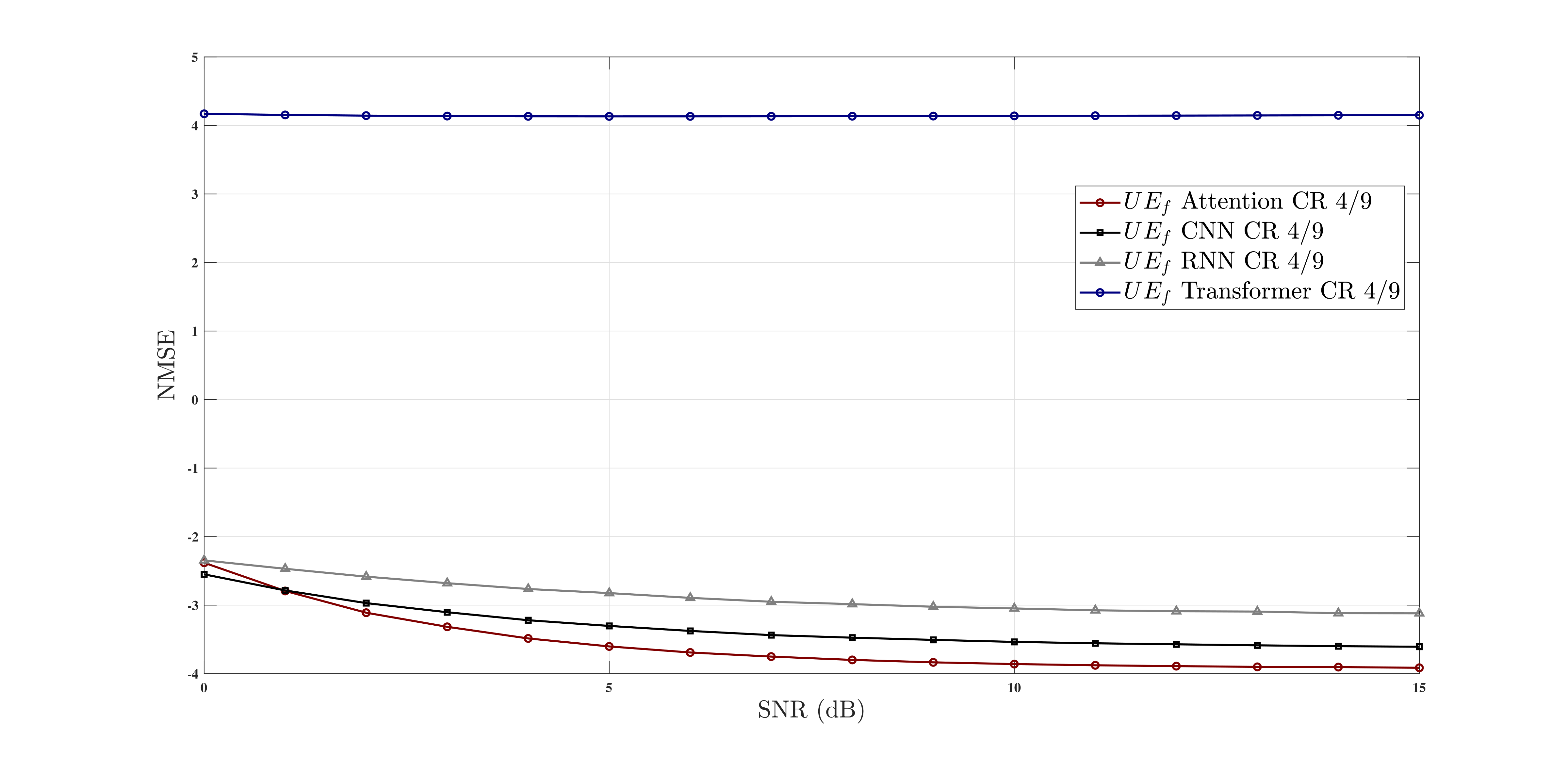}  
\caption{Comparison of NMSE across all employed architectures for $UE_f$ at a 4/9 compression ratio.}
\label{fig:nmse4tran_comp}  
\end{figure}

Comparing the loss results from the table \ref{tab:comp} and Fig.~\ref{fig:nmse4tran_comp}, it can be seen that the CNN+attention architecture outperforms all the other mechanisms deployed due to its ability to capture significant features to reconstruct the original input. Whereas, the transformer has the worst results out of all, making it an unsuitable compression method for this case. The CNNs and RNNs still showcase better performance than Transformers, but they fall slightly short of the performance achieved by CNNs combined with AM.

\section{Conclusion and Future Work}
This paper presents a novel integration of aerial RIS with CoMP-NOMA for next-generation multi-cell networks and investigates the feedback compression problem for the QPS in RIS-assisted wireless networks by utilizing Machine Learning autoencoder approaches. The proposed system demonstrates substantial improvements in spectral efficiency, reliability, and bandwidth efficiency, making it a promising solution for future wireless networks. Future research can explore the optimization of RIS placement and the development of advanced algorithms for real-time configuration of the RIS units. In addition to that, the impact of mobility and dynamic environments on the proposed system can be investigated.
















\bibliographystyle{ieeetr}
\bibliography{main}

\end{document}